\documentclass[defaultstyle,11pt]{thesis}

\usepackage{amssymb}		
\usepackage{graphicx}		
\usepackage{hyperref}		

\title{Social Media Analysis for Crisis Informatics in the Cloud}

\author{Mr.}{Gerard Casas Saez}

\otherdegrees{B.A., Universitat Politecnica de Catalunya, 2017}

\degree{Master of Science}		
	{M.S., Computer Science}		

\dept{Department of}			
	{Computer Science}		

\advisor{Prof.}				
	{Kenneth M. Anderson}			
	
\degreeyear{2019}

\reader{Prof.~Steve Voida}		

\abstract{  \OnePageChapter	

Social media analysis of disaster events is a critical task in crisis informatics research. It involves analyzing social media data generated during natural disasters, crisis events, or other mass convergence events. Due to the large data sets generated during these events, large scale software infrastructures need to be designed to analyze the data in a timely manner. Creating such infrastructures bring the need to maintain them and this becomes more difficult as these infrastructures grow larger and older. Maintenance costs are high since there is a need for queries to be handled quickly which require large amounts of computational resources to be available on demand 24 hours a day, seven days a week. In this thesis, I describe an alternative approach to designing a software infrastructure for analyzing unstructured data on the cloud while providing fast queries and with the reliability needed for crisis informatics research. Additionally, I discuss a new approach for a more reliable Twitter stream collection using container orchestrated systems. I finally compare this new infrastructure with existing crisis informatics software infrastructures and compare their reliability, scalability and extensibility with my approach and my prototype.}

\dedication[Dedication]{	
	To my family, my parents, my brother, my uncles, my grandfather who supported and encouraged me at all times on this adventure for the past three years.
	}

\acknowledgements{	\OnePageChapter	
	I would first like to thank my thesis advisor, Prof. Ken Anderson. Thank for his continuous support over the last three years. He gave me this amazing opportunity and welcomed me to join his group. I would like to give a special mention to the people that have helped me this last semester on the development of the vision I had for a new Project EPIC infrastructure: Sharan Srivatsa, Prashil Bhimani, Sahar Jambi and Ismael Garrido. Without their work, none of this would have been possible. I would like to also thank Afnan Meshal for pushing me to work harder and to think in new ways to design this infrastructure. To Rasha Mirza and Mazin Hakeem, I would like to thank them for their continued encouragement and support these past three years. Another special mention goes to Reem Albaghli for reminding me to take a break from time to time. I would also like to appreciate Jennings Anderson for giving wings to our crazy moonshots. I would like to thank the directors of Project EPIC, Profs. Leysia Palen and Ken Anderson for providing me with this amazing opportunity to conduct my research in their group. 
	
	I would like to thank my family for their support and encouragement on moving to the United States to work on the things I love doing, especially my mom for almost forcing me to not miss this opportunity. I would like to thank all my friends here in Boulder who have changed my life and given me wings to discover. Special thanks to my roommate Albert and the rest of my cohort of Balsells fellows: Roger Huerta, Noemi Collado, Jordi Ballesteros and Oscar Mu\~noz.
	
	Finally, I would like thank Pete Balsells, and all the team that makes the Balsells fellowship a reality, for providing me the opportunity to pursue my master's degree here in Boulder.

	}

\ToCisShort	

\LoFisShort	

\LoTisShort	

\begin{document}

\newtheorem{theorem}{Theorem}

\newcommand{\diff}[2]{\frac{\partial #1}{\partial #2}}
\newcommand{\diffr}[1]{\diff{#1}{r}}
\newcommand{\diffth}[1]{\diff{#1}{\theta}}
\newcommand{\diffz}[1]{\diff{#1}{z}}

\newcommand{\vth}{V_{\theta}}

\newcommand{\twochoices}[2]{\left\{ \begin{array}{lcc}
        \displaystyle #1 \\ \vspace{-10pt} \\
        \displaystyle #2 \end{array} \right. } 

\newcommand{\threechoices}[3]{\left\{ \begin{array}{lcc}
        #1 \\ #2 \\ #3 \end{array} \right. }    

\newcommand{\fourchoices}[4]{\left\{ \begin{array}{lcc}
        #1 \\ #2 \\ #3 \\ #4 \end{array} \right. }      

\newcommand{\twovec}[2]{\left(\begin{array}{c} #1 \\ #2 \end{array}\right)}
\newcommand{\threevec}[3]{\left(\begin{array}{c} #1 \\ #2 \\ #3 \end{array}\right)}
\newcommand{\twomatrix}[4]{\left(\begin{array}{cc} #1 & #2 \\ #3 & #4 \end{array}\right)}

\chapter{Introduction}
\label{introchap}
Our world is rapidly evolving and, due to the increased amount of people connected to the internet, we have seen a massive surge of data generated. Indeed, it is estimated that every month, seventy-two petabytes of information are moved around the Internet. This amount is expected to grow to 232 petabytes/month by 2021 \cite{ciscoreport}. This change forced software engineers to evolve and adapt and that, in turn, has led to work on an area known as big data software engineering.

NoSQL databases were created to respond to the need to store and analyze data at this new scale. These databases were simple engines that went back to basic storing and retrieval mechanisms for data. Compared to SQL databases that had been around for decades, these new NoSQL databases were really primitive when it came to analysis workloads. Over time, however, NoSQL has evolved to simplify its query syntax moving closer to SQL-like rule definitions. 
An example of this is Cassandra and CQL language which implements parts of SQL. 
Additionally, SQL engines have started to move towards supporting larger analysis workloads, slowly catching up to supporting NoSQL data sizes. This evolution is slowly resulting in better support for SQL queries on big data sized data sets. At the same time, however, the software infrastructure to make this all work has increased in complexity, making it more difficult to maintain these systems in the long term.

Furthermore, cloud platforms have begun to offer on-demand analysis tools. These tools are based on server-less environments which use ephemeral server instances to run their software. 
Examples of this are AWS Athena or Google Cloud BigQuery. 
This approach allows for the infrastructure to adapt its configuration based on usage and to scale up or down as needed. Thanks to improvements in the speed of on-demand computation, these cloud-based tools have become a great opportunity to cut costs and reduce maintenance difficulties for organizations wrestling with the demands of storing and analyzing truly large data sets.

One research area that benefits from work in big data software engineering is crisis informatics. Crisis informatics studies how the access to ubiquitous social media changes the way in which society responds to disasters. To perform effective research in crisis informatics, researchers must have access to tools that can collect and analyze large amounts of social media data collected during a disaster event. With respect to my work, the analysis of disaster data sets is an interesting problem domain. The frequency of queries performed over disaster data sets is low. There is no need for real-time, 24/7 analysis of these data sets. The project I am affiliated with---Project EPIC---instead collects data 24/7 and then analyses events later, sometimes months or even years after the event. Sometimes, the research questions around an event are known quickly and sometimes they take time to emerge. Given these conditions, I think it would be useful to see if the data analysis tools offered by cloud providers for big data can be usefully applied to the analysis of disaster data sets consisting of social media data. I think these offerings can be shown to be  both cost-effective and fast for domains with low-frequency analysis needs.

My thesis then involves designing and implementing an infrastructure for disaster analysis in the cloud.

\chapter{Background}

Before designing the infrastructure, I provide more information to help understand my problem domain. I present information on different topics. First, I talk about Project EPIC and its work on crisis informatics. After that, I present on how REST APIs work and how container orchestration technologis are valuable. Then, I dig deeper into cloud infrastructures, big data storage systems, cloud object storage, and cloud document analysis tools. After that, I talk about microservice frameworks and the differences between monolithic and microservice-based infrastructures. I also explain why I chose microservices for the design of my software infrastructure. 

\section{Project EPIC}

\begin{figure}[htbp]
	\caption{\label{fig:oldepicinfra}
	Project EPIC old software infrastructure diagram .
	}
    \begin{center}
	\includegraphics[width=100mm]{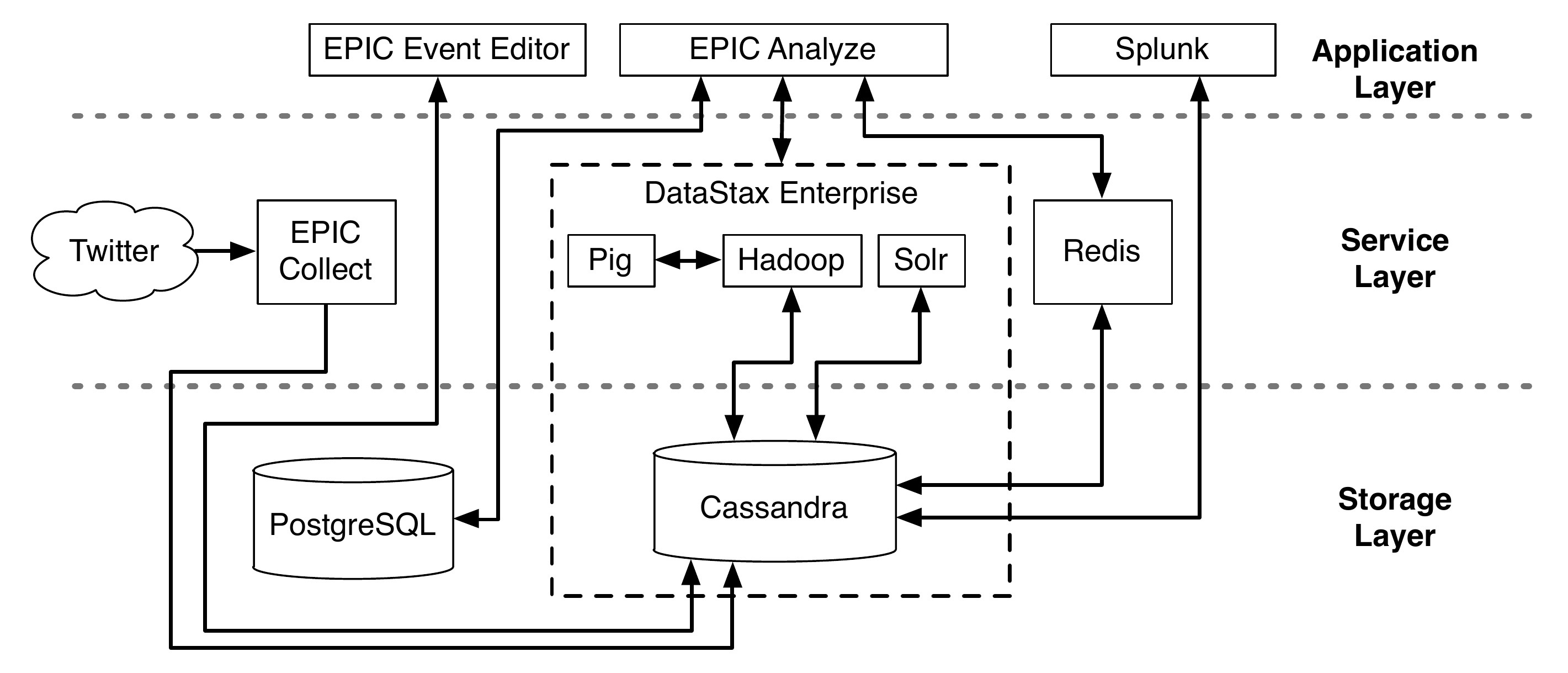}
    \end{center}
\label{xfigDiagram}
\end{figure}

Project EPIC (Empowering the Public with Information in Crisis) is a project at the University of Colorado Boulder. It conducts research on crisis informatics, an area of study that examines how members of the public make use of social media during times of mass emergency to make sense of a crisis event and to coordinate/collaborate around an event. Project EPIC has a long history of performing software engineering research. Since 2009, Project EPIC has been investigating the software architectures, tools, and techniques needed to produce reliable and scalable data-intensive software systems, aka big data software engineering\cite{Anderson:2015b}.

When large amounts of data are being collected, software engineers need to focus on structuring it so it is easy to perform analysis. They must work to ensure that the data is easily accessible to analysts. Project EPIC's big data software engineering research has explored these issues in depth during the creation of the previous Project EPIC software infrastructure that consisted of two major components: EPIC Collect and EPIC Analyze. EPIC Collect\cite{anderson2011design,schram2012mysql} is a 24/7 data collection system that connects to the Twitter Streaming API to collect tweets from various crisis events that need to be monitored in real time. Since 2012 to 2018, this software collected tweets with an uptime of 99\% and collected over two billion tweets across hundreds of crisis events. EPIC Collect used Cassandra as its storage layer. This NoSQL database is focused on writes and provided high throughput to handle all incoming tweets. EPIC Analyze\cite{anderson2015design,barrenechea2015getting} is a web-based system that makes use of a variety of software frameworks (e.g., Hadoop, Solr, and Redis) to provide various analysis and annotation services for analysts on the large data sets collected by EPIC Collect. In addition, Project EPIC maintained one machine---known as EPIC Analytics---with a large amount of physical memory to allow analysts to run memory-intensive processes over the collected data. 

The software architecture of EPIC Collect and EPIC Analyze is shown in Figure \ref{fig:oldepicinfra}. Note, this is a logical architecture that does not show how these systems are deployed. For instance, Cassandra was deployed on four machines that run separately from the machines that host the EPIC Collect software, Postgres, Redis, and the Ruby-on-Rails code that makes up EPIC Analyze. In all, the existing Project EPIC infrastructure was distributed across seven machines in a single data center maintained at the University of Colorado Boulder.

\section{REST API}

REST (Representational State Transfer) \cite{fielding2000architectural} is a well known architectural style that defines a set of constraints to be used when designing and implementing a web service. It recommends a pattern for making use of HTTP verbs on URLs that follow certain conventions for designing a natural way to design web services that then provide many of the same benefits provided by the Web itself. REST APIs provide an abstraction to a set of resources that shield the client from the internal representations of the data. Interaction is done through a HTTP request, making it easy to implement web applications thanks to its adoption across the internet. Designed as stateless operations, REST services help separate the user interfaces of clients from underlying technologies.

URIs are used to identify resources. That is, URIs make it possible to find an object in the system. Each URI must be unique for a single resource. A good practice is to format each resource URI in an ordered path structure. A lot of times a request to a service for a particular resource will return hyperlinks to other resources allowing a service to be built around a well-known URI and then letting clients discover other resources via service interactions. Each resource accepts a set of operations that derive from the HTTP operations: GET, HEAD, POST, PUT, PATCH, DELETE, CONNECT, OPTIONS. These can be used in pair with other parameters to execute a variety of actions on a resource. 

RESTful services allow for various CRUD (Create, Read, Update and Delete) operations to be performed easily. Here is a list of some CRUD operations mapped to an example REST service:

\begin{itemize}
    \item List all resources available \textit{GET /notes/}: Note the use of a plural with respect to the resource name as there are multiple notes in the folder. Also, use \textit{/} at the end, as it is functioning like a container or folder in a traditional desktop file system.

    \item Get a representation of a specific resource \textit{GET /notes/1}: This URI and HTTP verb will retrieve a unique representation for a single resource. We do not use \textit{/} at the end as we are accessing a specific resource.

    \item Create a new resource \textit{POST /notes/}: This should add a new element to the list of resources available. If no unique ID is provided, a new ID should be assigned by the web service and returned to the client. 

    \item Remove a specific resource \textit{DELETE /notes/1}: This should remove the specified resource from its containing resource. 

    \item Update a specific resource \textit{PATCH /notes/1}: This should replace the existing resource with the new representation that accompanies this request.

\end{itemize}

These operations can be augmented with parameters. We could add a GET parameter to filter down the results returned when we list all available resources. An example would be to list all notes that have not been archived. We could do so via ‘GET /notes/?archive=false’. This helps us cache future responses if the same filter is used again. These parameters are also useful when searching resources or for any other operations that may filter the result returned to the user. For creating and updating elements, we use the body from the request to include a representation of the object we are creating or updating. 

A common bad practice that should be avoided is using URIs for passing arguments. An example of such is adding the search term into the URI like \textit{GET /notes/search/hello}. This should be avoided since this is a parameter for a function, not a static resource to be retrieved or updated. Instead it should use GET parameters like ‘GET /notes/?search=hello’ as described before. 

For this project, I decided to use REST APIs to overlay an abstraction on top of my underlying infrastructure. Each API exposes a set of resources that allows the user to interact with the collection and analysis infrastructure. This abstraction layer allows changing the underlying infrastructure without affecting its clients. Also, thanks to the separation of data storage and user interface, I can change user interfaces individually independent of the underlying data. This separation allows having developers who are more front-end focused working on the user interface without the need to understand how the underlying system works.

\section{Container Orchestration Technologies}

As containerization technologies became more widely adopted---spurred by a recent migration to designing systems via microservices---large companies needed a way to manage their containers in a more friendly way. Interconnecting containers, managing their deployment, and scaling them to meet demand, were all possible, but were difficult to achieve. As a result, container orchestration systems were born.

To manage containers, these systems add an abstraction layer over containerisation technologies, making life cycle tasks related to containers (create container, launch container, pause container, etc.) easier to perform. There are a several container orchestration systems available at the moment. The most popular are Kubernetes and Apache Mesos.

In this project, I will make use of Kubernetes. The main reason for this is that---thanks to the open source community---it is easier to find tutorials and courses for Kubernetes. In addition, there are a lot of big companies backing this project and contributing to it; this activity provides evidence that this project will be supported well into the future. Finally, there is a lot of companies offering managed clusters on Kubernetes, which helps to migrate an infrastructure built on top if it if there was ever a need.

Google Cloud seems like the best fit to host Kubernetes as it has a managed cluster option that makes it straightforward to install and configure. In addition, thanks to Google being part of the maintenance team for Kubernetes, there is great support for Google Cloud services within Kubernetes.

\section{Cloud infrastructures}

Cloud infrastructures appeared in response to the need for big companies to maintain large sets of machines in their data centers for peak usage of their apps. The first company to create a cloud service division---and the one responsible for the popularization of cloud services---is Amazon. Amazon needed large data centers to be able to process the demand of the U.S. ``Black Friday'' holiday. However, during the rest of the year, they realized that their servers were mostly unused. They created the division of Amazon Web Services  to lease the computational power of those idle machines to external users.

Other cloud providers have appeared over the years. Google Cloud, Azure, and DigitalOcean are examples. With time, their offerings have switched from only leasing computational power to more complex tools. Some services offered by all these companies include managed databases, object storage, machine learning services, and more. Price and service availability is usually similar between them. 

With the rise of these services, software engineers have started to switch design practices to incorporate the cloud. Those architectures are designed to leverage cloud services for a more cost-effective system. Instead of using an in-house managed service they rely on cloud managed services for different pieces of their infrastructure. This allows savings on maintenance as the overall structure is simplified.

\section{Big Data Storage Systems}

For my thesis work, I will be collecting data from Twitter via the Streaming API. This API is limited to provide 1\% of the total tweets generated in Twitter every minute. Based on a report from 2013\cite{tweetsRecord}, I know that rate is approximately 5700 tweets per second on average. As a result, I can calculate the total number of tweets per second that I estimate will flow through my collection infrastructure. That is approximately 57 tweets per second on average as a minimum bound. Since that corresponds to 4.9M tweets/day, I need a data storage technology that scales to handle large data sets. 

Previous studies in Project EPIC pointed to the usage of NoSQL database. However, systems like this can become a huge bottleneck for analysis workloads. Collection storage needs to be separated from analysis storage in order to increase reliability and avoid analysis queries causing a computational bottleneck that can affect the overall performance of the collection pipeline. 

In addition, I believe it is important for the future to keep data in the raw format in which it was received. Storing raw data will allow the evaluation of new tools. In addition, storing unstructured data as documents is better than converting the original format into a specific database representation since Twitter can (and does) change its data structures at any time.  

\section{Cloud Object Storage}

Most cloud service providers offer object storage services, also known as blob storage services. These services are documents stores abstracted as file systems. They provide cheap and reliable file storage. In addition, they provide programmatic ways to add files, download, and list them from code. Cost is based on the total storage needed, which avoids having to pay for extra disk space when its not being used. At the same time, it functions as an abstraction on top of specific disk configurations to simplify management.

Thanks to the file system structure, folders can be used to provide fast filtering by key. This provides guidance in how data should be structured. A clear example of a use case would be to store all tweets collected during a day in a single folder. This would allow us to filter tweets by date quickly. Such keys can be used as a hash table index with prefix look ups. 

Each object stored has internal metadata that works as a means of deciding how to store the file internally as well as how it should be accessed externally. Part of the metadata is used to decide where it should be stored. Files that are accessed rarely can be stored in cold storage for a cheap price. Files that need to be downloaded frequently are usually stored in fast SSD disks to reduce latency. Other options include replicated files across regions allowing the ability to serve files from the closest region to a client. 

Given that in disaster analysis we tend to explore tweets for events individually, individual data elements can sit for a long time without being accessed. This characteristic makes this data perfect for cold storage as it allows to keep costs low while making sure we do not lose any of the data. In addition, having an interface to access the data whenever we really need it, makes this option worth it. 

\section{Cloud Document Analysis Tools} 

Due to the rise of unstructured data, there has been a lot of work within the data analysis world to bring SQL queries to large data sets. Various NoSQL technologies brought parts of SQL to life by providing abstractions on top of their own data access systems. The main issue with those systems is that it needed to have two separate machine clusters, one for collection and one for analysis workloads to ensure the two functions do not overlap with each other or compete for resources. In addition, storage can make use of different formats across the two tasks, which makes it difficult to switch data analysis tools on top of an underlying collection without having to migrate the whole data set to a new format. This can be a problem if a product used is discontinued or slowly abandoned over time.

To address that issue, different alternatives have been created. Hive on top of Hadoop was an attempt of bringing SQL to the MapReduce world. A more recent alternative has been Presto, an interactive query system that can operate quickly at petabyte scale. Created inside Facebook, it is similar to Hive while operating primarily in memory. Its main goal is to allow for interactive queries (i.e. fast response times) to be performed on top of unstructured large data sets. 

Given that analysis workloads are sparse, adapting Presto to use server-less architecture on the cloud makes sense as it can help keep costs low. This allows for computing resources to be created when needed to resolve a query and also allows for high parallelism to reduce the time to compute the result of a query. Cloud services similar to Presto include Athena from Amazon Web Services and BigQuery from Google Cloud. They both offer similar products on a pay-per-data-element-analyzed basis and can be pointed to their own object storage services for data access.

Making use of this type of service, I can keep simple structures of files in object storage while allowing for quick yet complex analysis workloads that are run in parallel without having to support a large infrastructure to do so. This solution can bring great value, especially in fields like disaster analysis where data is not being analyzed continuously. 

An example query that I can do in this system would be to get the number of tweets for each user in a data set and order the result. Using Google BigQuery, I can run this query with a simple SQL query on top of a zipped set of JSON files stored in Google Cloud Storage. 

\begin{table}[htb]
    \caption[Table comparing execution times to data set size]{
    Table comparing query execution times for the top 50 most active users in an event with the size of the overall data set.
	}
    \begin{center}
    \begin{tabular}{|r|r|r|} \hline
	 \textbf{Storage (Gb)} & \textbf{Number of tweets} & \textbf{Query Time (seconds)} \\ \hline 
	
0.03 & 5638 & 2.2 \\ \hline
0.472 & 72027 & 4.1 \\ \hline
2.5 & 408422 & 5.3 \\ \hline
23.1 & 3776311 & 25.2 \\ \hline
204.2 & 34356642 & 111 \\ \hline

	\end{tabular}
   
\end{center}
\label{powertable}
\end{table}

\section{Microservice Architecture}

Microservices is an approach to distributed systems that promote the use of small services with specific responsibilities that collaborate between them, rather than making use of big components with a lot of responsibilities that make interactions more difficult. They are thus more cohesive units of software with minimal dependencies between them. A system designed with loosely-coupled, highly-cohesive software components has always been a highly desirable goal in software design, and microservices help to achieve that goal within distributed software systems\cite{gradybook}.

Compared to monoliths---large software projects with a lot of responsibilities---microservices allow for faster iteration cycles and deployment. This is achieved thanks to the code base for each service being separated. Due to that, developers can works on parts of the bigger system while not overlapping with others. This flexibility also reduces overall system complexity \cite{microservices}. Thanks to the popularization of container orchestrated systems, microservice architectures have become easier to manage and deploy. Container orchestrated systems provide an abstraction that matches really well with microservices. 

Another advantage of microservices architecture is that it allows for different frameworks and languages to be composed into a single system architecture. Even though this can be interesting to accommodate developers, it is also acknowledged that having a really diverse code base can make it complex for a developer to jump between parts of the system. For that reason, it is recommended to maintain a single language and framework when it comes to designing a microservice architecture. It is still interesting to have the option to jump to a new language or framework to use custom features only available within them. An example would be having to create a stateless service that handles a large number of messages. It would be interesting to rely on frameworks and languages focused on parallel processing and message passing to improve performance. For the decision of what language should be the main framework for a given microservice-based architecture, developer adoption, reliability, and ease-of-use should be the principal criteria.  

Finally, microservice architectures allow for different components to be updated over time and optimized independently. This can benefit the whole infrastructure by helping mitigate bottlenecks on a system individually. In addition, given that the structure is loosely coupled, it allows for services to switch their internal implementation while keeping the same external functionality. An example would be to change the database used by a specific service. The service could be replaced fully by itself without affecting the rest of the system. This simplifies the contracts between teams of developers and reduces dependencies to the APIs of other microservices. 

Having small microservices do specific tasks makes development cycles faster and more independent. It also makes incremental deployment much easier and less dangerous. Finally, microservices make it easier to scale software systems since one can individually scale parts of the system depending on their usage

Most languages have frameworks that allow for easy REST microservices implementations to be created. Dynamic languages frameworks work great for prototyping and allow for fast development. However, they can become difficult to manage unless you make sure your code base is extensively tested. Any line of code can fail due to dynamic typing at execution time and this is dangerous in the long term. This makes it dangerous for production systems as you lose the advantage of analyzing the code when compiling. On the other hand, statically compiled languages make for a more reliable code base as it checks types at compilation time. This can make the code more stable and help detect errors on the code base earlier in the process. 

An example of a framework for microservices would be Dropwizard. This framework, based on Java, provides a complete reliable backbone for microservice development. It includes several libraries to handle database connections and other service connections. Its design is also simple enough to work with so that the learning curve is low.

\chapter{Problem statement}
\OnePageChapter
\label{problemstatement}

The goal of my thesis is to explore and compare the benefits and limitations of using cloud services to implement disaster analysis infrastructures. After the existing Project EPIC infrastructure went down, it was very important to design and implement a new infrastructure that could be more reliably maintained. The old infrastructure was manually deployed on a set of physical machines in a local data center and was not developed using microservices or containerization. My hypothesis is that I will be able to achieve greater reliability, extensibility, and scalability of a system with a new cloud-first infrastructure with significantly-reduced maintenance costs. I will also explore whether I can improve the ability for teams to work on the infrastructure simultaneously.  My specific research questions are:

\begin{enumerate}
    \item Does the new infrastructure have lower maintenance costs than the existing infrastructure? 
    \begin{enumerate}
        \item Is it easier to scale? 
        \item Is it easier to extend? 
        \item Is it more resilient to failures? If so, how?
    \end{enumerate}
    \item Is it easier to perform parallel development to extend the infrastructure than the previous infrastructure?
\end{enumerate}

\chapter{System Requirements}

To focus my thesis and make it easy to evaluate, I present the requirements for my prototype. 

Disaster analysis systems need to be able to collect social media data for various events such as cyclones, floods, earthquakes, and hurricanes, whenever they occur using the live stream of data from social media platforms. The collection needs to be reliable and capable of keeping up with peak message rates whenever needed. The system needs to be running continuously and should be able to collect data for multiple events in parallel. There should be an interface to manage events collected and to start new event collections.

Due to social media streams not having static data schema, data should be saved without any modification in an unstructured data storage mechanism. This storage needs to support fast retrieval of data elements for a single event. In addition, this solution should allow keeping the data for long periods of time with low usage so that future analysis can be run retrospectively on all historical data collected. The data storage should have a reliable mechanism in place to avoid losing any data successfully stored.

Data should be accessible by event via an easy-to-use user interface. This interface should allow analysts to scan the data quickly and be able to explore in depth if needed. In addition, analysts should be able to time slice the data to better understand what was happening at different times of the collection. These characteristics paired with a good data visualization of the number of messages collected by hour should help analysts explore user interactions throughout the time line of the event. Being able to understand in depth what was happening during an event in a timely manner is important as it helps analysts study the behavior of members of the public responding to or talking about a disaster event.

Finally, the system should allow for interactive queries to be performed in a high-level language like SQL. These queries should be computed with low latency to enable fast data exploration. This feature can help researchers better understand the data they are working with, enabling them to explore the data set in multiple ways. This exploration could enable better insight into how people react to emergency response. At the same time, it could become a way for first responders to better understand a situation with flexible knowledge of what is happening around them via social media messages. Reducing latency could prove really useful in managing interactions with the public in real time.

\chapter{Approach}

In this section, I present the design of my replacement for the original Project EPIC infrastructure. As mentioned above, my design relies on a microservices architecture that is deployed on a cloud-based infrastructure making use of container orchestrated technologies to coordinate all services and deploy them in a distributed fashion. Thanks to this design, I am able to use fewer physical machines while optimizing the computational power I have available.

For the system description, I will use a comparison with the previous existing Project EPIC infrastructure, comparing each component with the new proposed infrastructure. First, I will discuss the collection architecture and its persistence layer. Then I will describe the analysis infrastructure and I will end by describing my user interface layer and comparing it with EPIC Analyze.

\section{Data Collection and Persistence}

\subsection{Previous infrastructure}

\begin{figure}[htbp]
	\caption{\label{fig:oldinfrcollection}
	Diagram highlighting the data collection parts in the old infrastructure.
	}
    \begin{center}
	\includegraphics[width=150mm]{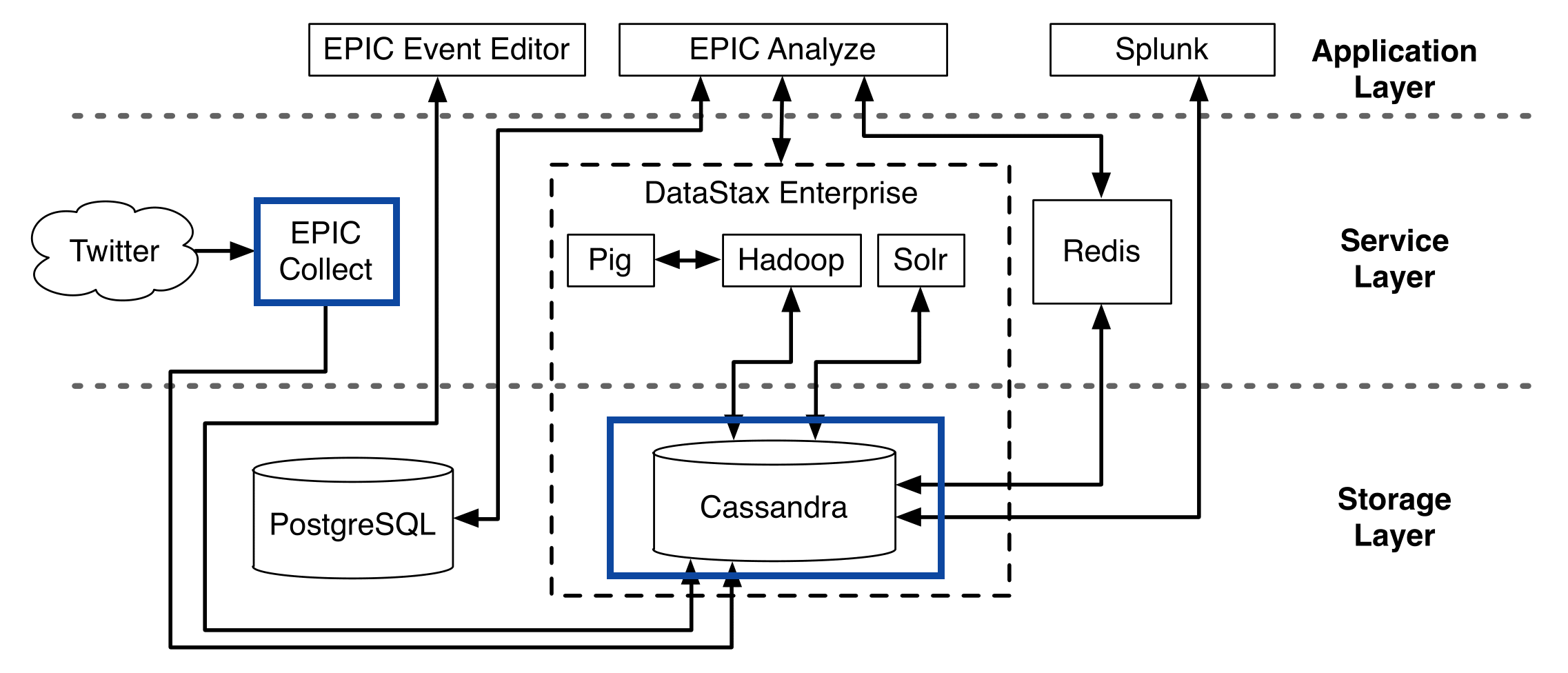}
    \end{center}
\label{xfigDiagram}
\end{figure}

Data collection in the previous infrastructure is mostly composed by the monolithic EPIC Collect. The system is designed to be composable within itself by using the Spring framework to do dependency injection between different services. To collect tweets, it connects to Twitter’s Streaming API directly. It uses a separate thread to monitor the keywords needed and restarts the collection thread when the keywords are updated. Those keywords are updated through the EPIC Event Editor. This web application allows users to create and stop events and their associated keywords. Updates via this web application changes fields in the EPIC Collect database; these changes are then detected by the EPIC Collect thread which restarts its collection to make use of the new keywords. 

To process the tweets and separate them into events, the system keeps an in-memory queue that separates the ingestion part of the infrastructure from the classification component. The classification mechanism checks events tracked at the moment and classifies tweets to the events it matches. This process, once finished, sends tweets to the persistence layer. 

Data was stored first in MySQL and then in Cassandra. Project EPIC researchers were forced to switch to Cassandra after realizing that using MySQL created a bottleneck when storing tweets. In addition, to process data faster, there is a need for data to be distributed across multiple machines to increase parallelization and allow real-time computation. Data is stored using the event id as the key. This allows retrieving tweets by event faster. In addition, to allow for quick date range queries, it includes the day in the key such that time-slicing queries are fast. The design of the key also helps to balance data evenly across the Cassandra cluster \cite{anderson2015design}. 

\subsection{New infrastructure}

\begin{figure}[htbp]
	\caption{\label{fig:newinfracollection}
	Diagram highlighting the data collection parts in the new infrastructure.
	}
    \begin{center}
	\includegraphics[width=150mm]{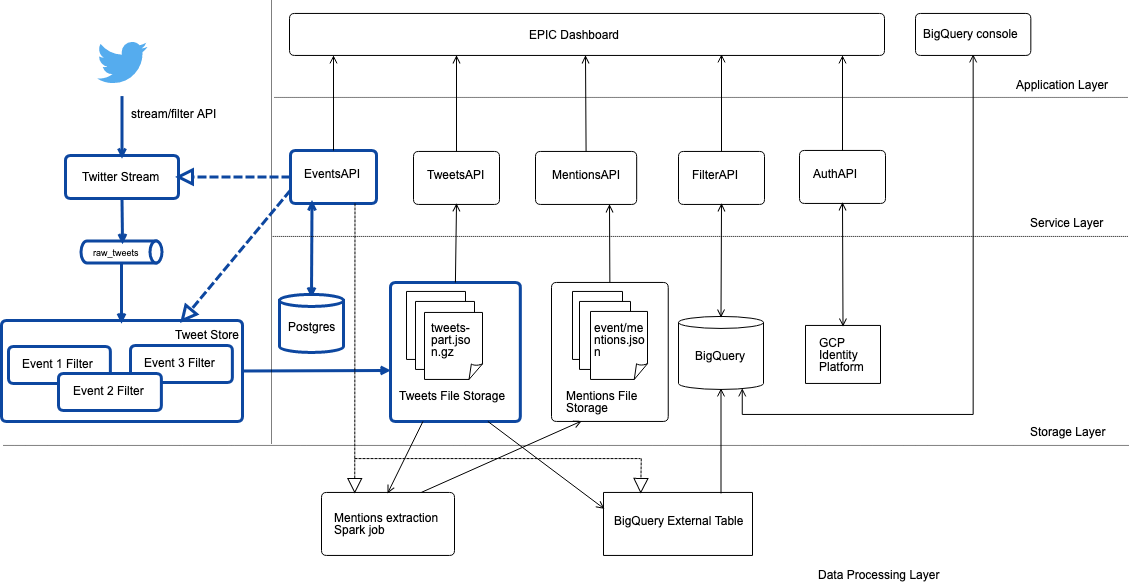}
    \end{center}
\label{xfigDiagram}
\end{figure}

In order to make collection faster, I designed the collection components to be stateless. That is, these components are designed to perform operations based only on the messages they receive. This makes the system more reliable as it allows the underlying container orchestration system to replace crashed services and scale up if needed. 

Data collection from the API needs to be really reliable. It also needs to be available to ingest multiple events at the same time. One of the main concerns with this part of the infrastructure was how to avoid losing messages at peak times while avoiding having a huge infrastructure running all the time. For this, I designed the ingestion pipeline to use Kafka between stages so that it can cache messages during peak times and process them all at once after the rate of messages decreases. It also acts as another reliable mechanism that can help crashed services recover non-processed messages once they are back up. This acts similarly to the in-memory queue from the old infrastructure but is more reliable. I separated the pipeline into two services: ingestion and filtering.

For ingestion, there is the Twitter stream connector. Its function is to connect to the Twitter Streaming API using a defined set of keywords and send incoming messages to Kafka. The set of keywords are pulled from a Configuration Map in Kubernetes. This service is periodically checking a local file with the keywords to restart the collection if they change. Thanks to the microservice approach, I was able to leverage the Go language for this microservice instead of Java. This is because I needed increased reliability and good message processing capabilities. Go provides a good network interface and allows for quick message processing. Note that this service does not process tweets in any way, it only retrieves them and sends them down the pipeline to Kafka. 

The next step in the pipeline is the filter service. This service is fully stateless, it relies only on the messages in Kafka. In addition, each service instance is event specific. The system only runs an instance of the filter service for each event active at any given time.  If no active event is being tracked, there are no filter instances running, which means that it is only using computing resources when needed, allowing us to use more resources for analysis. 

For reliability, when an instance goes down, it reconnects to the latest offset committed to Kafka. I use consumer groups from Kafka to distinguish between different filter services reading at the same time from the same topic. Once a tweet arrives at the filter service,  the service decides whether or not it belongs to the current event. It does so by doing a simple string search on the tweet JSON. If any of the keywords from the event are in the JSON, that tweet is classified as pertaining to the event and buffered. 

To store tweets, I rely on a cloud object storage service. This service serves as an online file system, abstracting away data distribution and replication. This allows me to work under the contract of reliability from the cloud provider and forget about implementing reliability measures like data replication. In addition, thanks to the availability of long term storage options, I can keep data storage costs low. I have programmed the data to fall into cold storage after a year, allowing the infrastructure to retrieve the data faster while it is being collected, and reduce costs after a year has passed since the need for that data is likely greatly reduced at that point.

The filter service is in charge of uploading tweets to the object storage service. Every hour, the service compresses all messages in its buffer and uploads them to the object storage service as a single file. To handle peak usage times while ensuring good performance for analysis, the service will also upload messages if the buffer fills to 1000 tweets. 

To provide fast querying of tweets given an event and to provide support for time slicing queries, I use the file system abstraction to store  metadata on the filename and path. On one hand, I store each file into their corresponding event folder. This allows listing all files with tweets for a specific event fast, as it is only listing all files under a folder. On the other hand, I store the date and the number of tweets in a file in its filename. Thanks to this design, I can later parse the filename and detect if I have any tweets that I need in the file. File names thus work similarly to an index. I keep the number of files low by setting the buffer size limit high. Thanks to this index, I also can create a visualization of ingestion rate per hour without needing to access the files. 

Organizing all this, I developed the Events API. This service is in charge of performing CRUD operations to events. It uses a PostgreSQL database to store the events data and their status. It also keeps track of when an event collection is started or stopped, and which user is responsible for the action. This service is also in charge of updating the pipeline described above if needed. This service interacts with Kubernetes to instantiate new filter services when an event is created or a collection is restarted. It also manages the configuration map that maintains the set of keywords to be collected at any given point in time. This service is in charge of abstracting all event-related actions. It exposes a REST interface used by the infrastructure's user interface.

\section{Data Analysis}

\subsection{Previous infrastructure}

\begin{figure}[htbp]
	\caption{\label{fig:oldinfaranalysis}
	Diagram highlighting the data analysis parts in the old infrastructure.
	}
    \begin{center}
	\includegraphics[width=150mm]{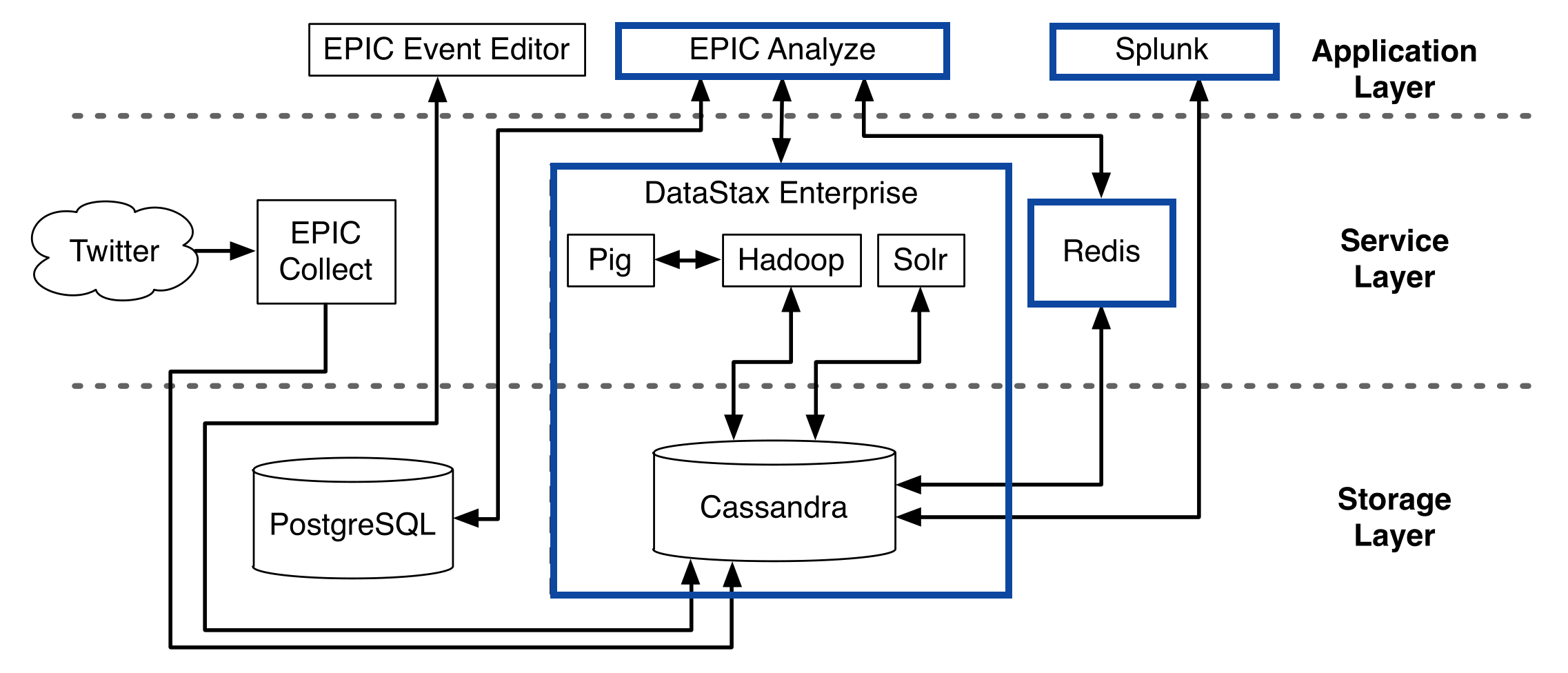}
    \end{center}
\label{xfigDiagram}
\end{figure}

Previous work on Project EPIC on data analysis is presented across several papers. First most work started as a SQL query when data was still stored in MySQL. Due to the switch to NoSQL technologies for storage, the technical difficulty to access collected data increased. This, in turn, made it complicated to filter data sets and explore them. To address this situation, software engineers had to interact with analysts to narrow down data sets and expose them in a format that would be easy for them to import into third-party tools. This was a highly manual process. 

To improve that process, EPIC Analyze was created. This monolithic system used a model of data sets associated with events to load and visualize the data. Due to Cassandra not being good at supporting random accesses and the main cluster needing to be available for peak ingestion times, the analysis needed to happen in a separate cluster. EPIC Analyze had its own Cassandra cluster where data sets to be analyzed were stored. The process of loading the data to be analyzed was done manually. The main functionality available in EPIC Analyze was browsing and pagination, filtering, and the job framework.

For browsing and pagination, EPIC Analyze allowed exploring the data set at the tweet granularity level. It paginated the data using Redis indexes created when data was imported to the system. Each entry in the index pointed at the column and key where a tweet was in Cassandra. Thanks to Redis, pagination was fast. This allowed analysts to interact with each other by finding important tweets and sharing pages where interesting tweets were located.

Filtering was aimed at reducing the size of the data to be explored. EPIC Analyze allowed analysts to specify a search query based on various fields from a tweet. This took advantage of the DataStax Enterprise integrated version of Apache Solr with Cassandra. Solr built indexes when data was loaded into the analysis cluster, allowing for sub-second queries when the index was finished.

Finally, for other analysis techniques, EPIC Analyze had a job framework available to extend the capabilities. This framework used Resque to queue jobs and perform them. Some of those jobs used the Apache Pig QL language to query the dataset for deeper insights. Output results were stored in Cassandra in the form of JSON. This was accessed by EPIC Analyze and showed in the UI either in RAW form or in visualizations if implemented.

A limitation on this design is the fact that data needed to be transferred to a separate cluster for analysis; this made the job of analyzing the data set asynchronous to the event itself. This can be dangerous, as analysts are blind to what is happening in an event until the event is loaded into EPIC Analyze. In addition, even though Datastax does a great job at integrating Hadoop with Cassandra, performance is hugely influenced by what is happening in the Cassandra cluster. This means that two MapReduce jobs running on separate clusters will find a bottleneck when they try to read from Cassandra at the same time, as the cluster will need to perform both reads simultaneously. This is similar to an issue I encountered with the infrastructure proposed in my undergraduate thesis. There Spark had issues loading data to memory while Cassandra was collecting data since Cassandra was running out of memory.

\subsection{New infrastructure}

\begin{figure}[htbp]
	\caption{\label{fig:newinfraanalysis}
	Diagram highlighting the data analysis parts in the new infrastructure.
	}
    \begin{center}
	\includegraphics[width=150mm]{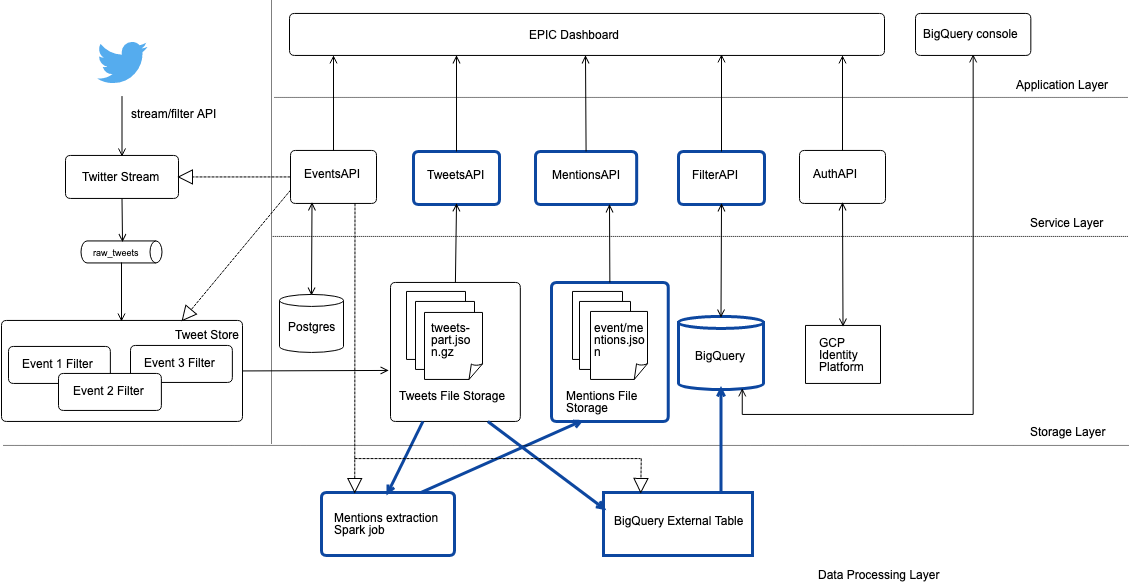}
    \end{center}
\label{xfigDiagram}
\end{figure}

As previously discussed, there is a need to explore and analyze collected disaster data sets in a way that analysts can understand what people were discussing during the event. On one hand, my infrastructure needs to allow users to scroll through an entire data set with ease, abstracting the underlying storage layer. On the other hand, I want to provide a way to query the data set with a high-level language like SQL for interactive exploration of the data set. This would allow analysts to test theories and understand the data set at a deeper level than what browsing provides. In addition, my system needs to be able to do some analysis in batch mode after finishing a collection to help load an event's data more quickly into the user interface.

For data exploration, I built the Tweets API. This microservice abstracts the underlying data storage layer. It uses the filenames with tweet counts to create a pagination index. This index is used to calculate what file is needed to retrieve particular tweets. This file is calculated from the index given a page number and the number of tweets per page. This index can be created given a list of all the files under an event's directory. Due to Google Cloud Storage API’s being slow when the number of files to list starts to exceed a certain size, I decided to cache the index in the memory of the Tweets API microservice for 10 minutes. This allows analysts to quickly explore data sets, being able to load any page in seconds. Thanks to this API, I abstract away the storage layer from the user interface. In addition, as the event name is passed as a variable, this API is completely independent of the Events API. This follows the software engineering rule of reducing dependencies as much as possible. In addition, it also allows for performance patches to be deployed on the exploration part of the infrastructure independently from the collection-related components.

Similar to EPIC Analyze, I also added the ability to annotate tweets. Analysts can add arbitrary tags to tweets in a data set. These tags are color encoded using a hash function based on the text to enable fast visual exploration. These annotations are stored in PostgreSQL using the Events API.

Another factor that takes importance is time slicing during data exploration. Using the date on the filename, I can filter what part of the pagination index I want to use. This can be done in linear time. After the filtering has happened, I can calculate again what file to retrieve.

For filtering, I rely on Google Cloud BigQuery service. This service allows me to query the dataset in SQL interactively. I use the Filter API to abstract away the details of the interaction with BigQuery. This API returns data in a format similar to the Tweets API. The difference between the two is that the data returned is only the data needed for presenting an overview of a tweet. I did not implement all the filtering options that existed in EPIC Analyze, but I do know that the same functionality can be achieved over time. At the moment, I implemented only text search using the like operator on BigQuery. I use temporal tables to paginate through the results. To access details from a tweet, I provide the filename and tweet id to the Tweet API. That API can then retrieve the file and return the full tweet JSON to the user interface.

Thanks to BigQuery, I can also personalize and export data into CSV format on demand. Using SQL, I can narrow down the data set and choose the fields I want to expose. Once that is done, I can export the results into a CSV stored in Google Drive. This is especially useful for Project EPIC interactions with external collaborations.

For additional analysis jobs, I leverage Google Cloud Dataproc and Spark. I created a workflow template on Dataproc which gets triggered every time an event collection is stopped by the Events API. This workflow can be modified independently of the rest of the infrastructure by adding new Spark jobs. Jobs receive the event name as a parameter. Spark jobs are written in Java and stored in Google Cloud Storage as jar files. This jar is loaded on a cluster created when the workflow is triggered. Creating clusters on demand avoids having unnecessary servers sitting around when there are no analysis jobs running.

The first Spark job added is the mentions extraction job. I believe that extracting the most mentioned users from a data set can lead us to find the most important users in a data set. This information can be later used to extract their timelines and contextualize the collection further. 

Developers must also create APIs to expose the data that Spark jobs extract. In this case, there is a Mentions API that knows how to access the resulting output from Spark and serve it in a paginated fashion. 

Note that adding a new batch job does not need any interaction with the rest of the services. This allows for new developers to work on extensions without any need to change existing services. The only step needed to interact with other code is the user interface; I will that component in more detail below. In the mentions extraction case, to prove that this was a good approach for parallel development, it was developed on the side of the rest of the infrastructure by a different developer. This developer worked by himself to create this extension. The only part of the code base that he needed to change in my system was within the user interface. This provides a powerful way to extend the capabilities of the system in the future.

\section{User Interface}

\subsection{Previous infrastructure}

\begin{figure}[htbp]
	\caption{\label{fig:oldinfrui}
	Diagram highlighting the user interface parts in the old infrastructure.
	}
    \begin{center}
	\includegraphics[width=150mm]{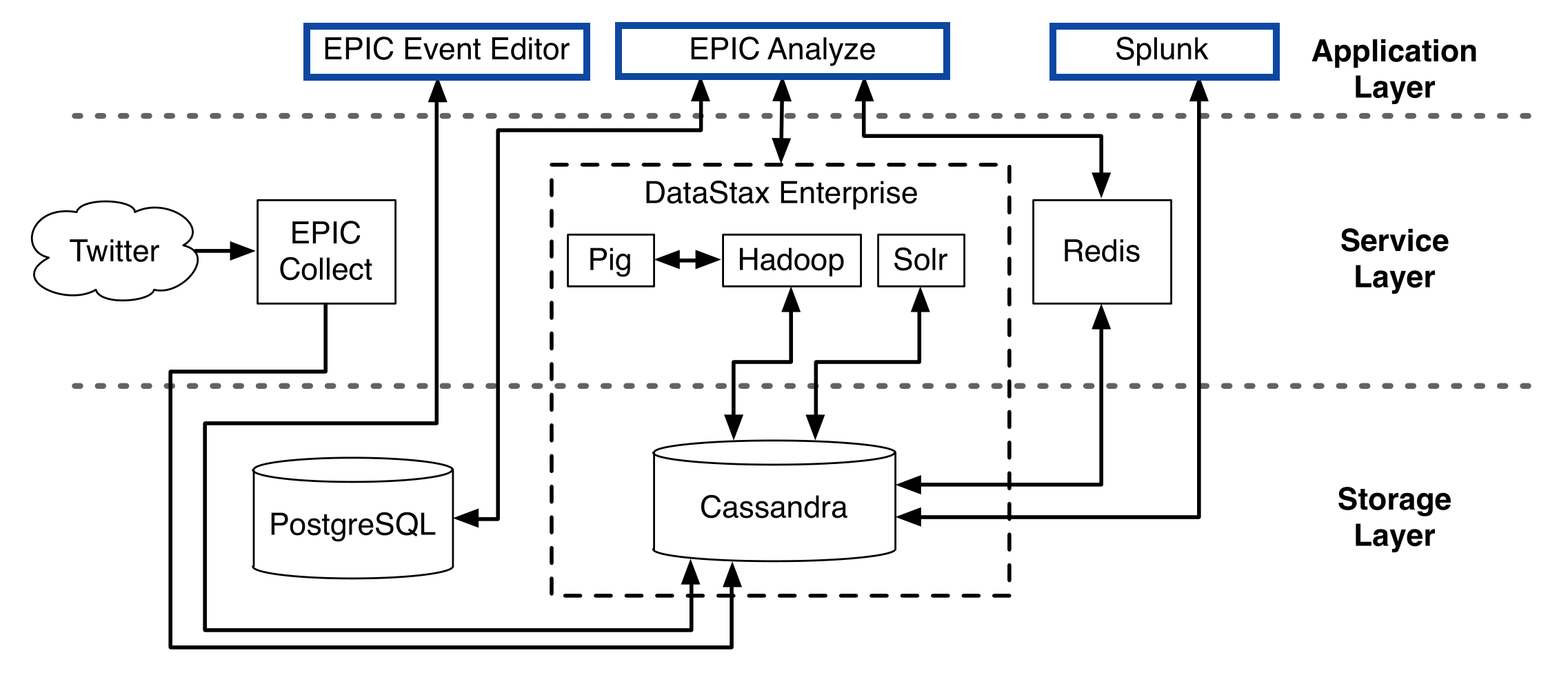}
    \end{center}
\label{xfigDiagram}
\end{figure}

There have been several iterations on the user interface for EPIC. The latest was developed as part of EPIC Analyze. This monolithic app is built on Ruby on Rails. Coupled with Redis and Cassandra, it provided an interface for data set interactive exploration. The interface was done in the form of a web app. Designed with Ruby on Rails views, the code was highly coupled to the underlying infrastructure. This meant that to add new capabilities to the existing user interface, developers needed to expand on the already existing application, creating a highly coupled system that increased in complexity over time. 

Focusing on the web application, we can find a list and detail view to explore tweets in a data set. Tweets can be explored page by page in the browser, and each tweet can be expanded to see all of its related metadata. On top of the page there is a timeline that shows the volume of tweets over time for the existing data set. The analyst is allowed to select a period of time by drag-and-drop interaction on the timeline, allowing for the data set to be time sliced. EPIC Analyze also provides a querying interface to perform complex queries with composed conditions. This functionality allows for analysts to filter down datasets by different elements from the tweet metadata. These queries also update the count timeline to reflect the tweet volume of the new filtered data set.

Another interface element provided was the EPIC Event Editor. This was a simple user interface to associate keywords to events. This was plugged in to EPIC Collect to determine what tweets to retrieve for each event. From this interface you could start and stop collections at any given time. This system was completely independent from EPIC Analyze, as the later needed data to be loaded into manually to analyze it as discussed above. This requirement was due to the design of the previous infrastructure. 

Finally, the last element of the old user interface is Splunk. This is an application that manages and allows for queries to be written in a domain specific language on top of Cassandra. This allowed for real time queries. The main problem was that you needed to use the Splunk pipeline language to perform queries. Splunk was also completely independent from EPIC Analyze and the EPIC Event Editor interfaces.

On the user management aspect, EPIC Analyze exposed a set of abstractions to manage permissions and access to data sets. User affiliations restrained access to certain data sets. User accounts needed to be created on the Postgres database to support this. Splunk had its own authorization mechanisms as did the EPIC Event Editor, all independent from each other. 

\subsection{New infrastructure}

\begin{figure}[htbp]
	\caption{\label{fig:newinfrui}
	Diagram highlighting the user interface parts in the new infrastructure.
	}
    \begin{center}
	\includegraphics[width=150mm]{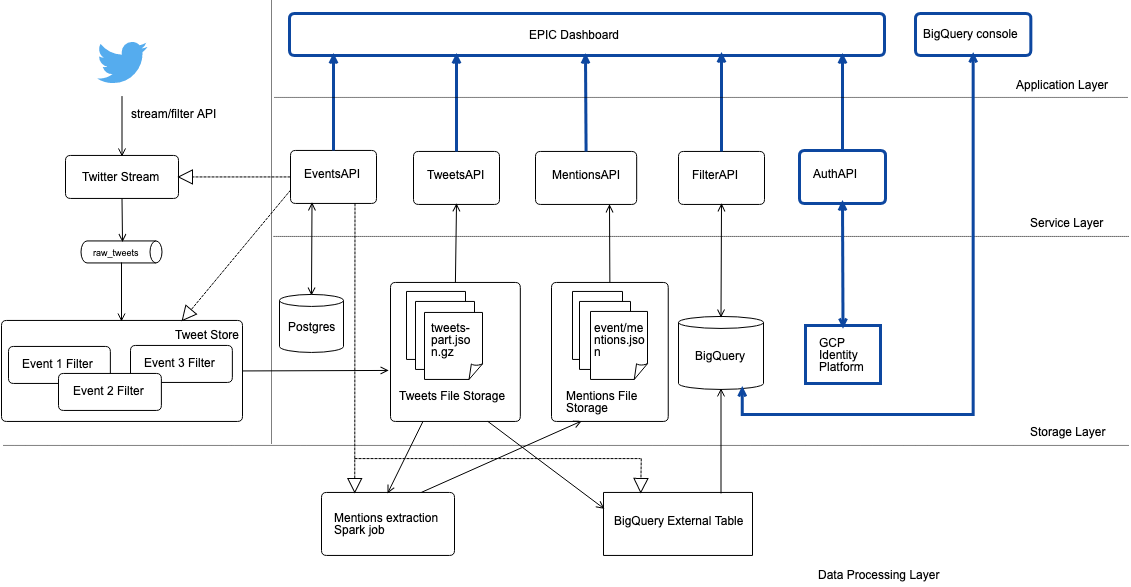}
    \end{center}
\label{xfigDiagram}
\end{figure}

For the new infrastructure, I looked for a front-end framework that would be able to modularize each user interface element independently. The main reason behind this decision was to allow for various developers to work on interfaces of their own APIs independently. There are many frameworks available to develop front end applications. My goal was to find one which is suitable for building real-time dashboards. 

According to a study, the most popular front end frameworks are React JS, Vue JS, and Angular JS. Upon more research, in 2019, 78.1\% of front end developers use react, 0.8\% use Vue and 21\% use Angular. The clear winner in terms of frameworks that I wanted to use was React JS. React JS also breaks down its pages into components, each of which can be developed in parallel. The way React works is that when a component receives new data, only part of the page refreshes  which is exactly what I needed for my infrastructure's user interface. Each API in our dashboard can be treated as a smaller independent application. This separation needed a framework for application state management, that is a tool which is responsible to segregate functions and data. The best application framework out there to currently work seamlessly with React is Redux.

This approach to our user interface design allows teams to work in parallel on different elements from the user interface and compose it all on a single application app. With this approach, an API developer, can decide on their own data visualizations and integrate them into the rest of the user interface separately. In addition, the interface is more robust as it is fully independent of the back-end code and it only interacts with the JSON data representations exposed by the various APIs.

\begin{figure}[htbp]
	\caption{\label{fig:mentions}
	Most mentioned users in a dataset, ordered by number of mentions descending.
	}
    \begin{center}
	\includegraphics[width=150mm]{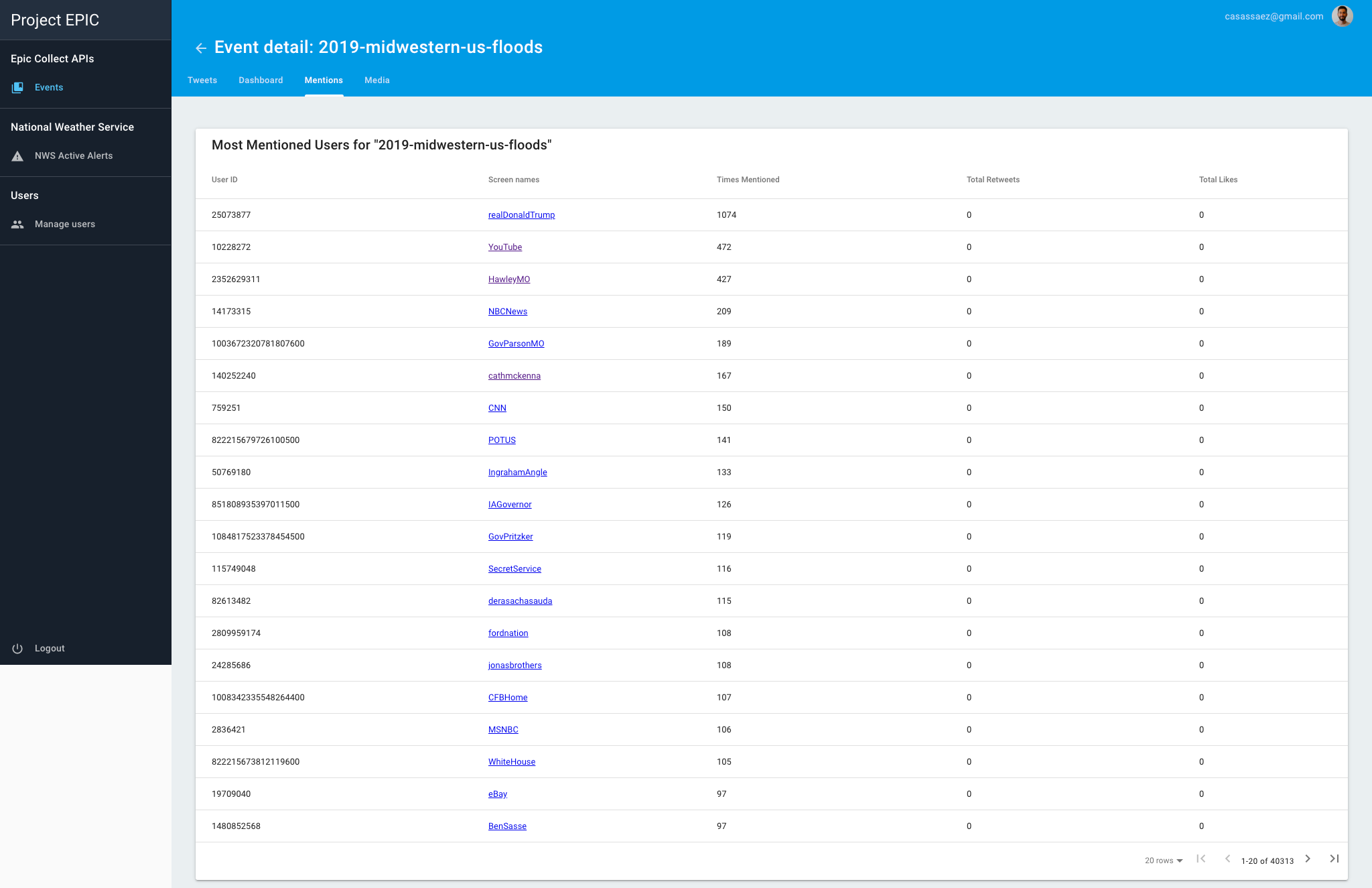}
    \end{center}
\label{xfigDiagram}
\end{figure}

An example of such a component is the Mentions integration (see Figure \ref{fig:mentions}). It integrates the mentions from the events in a new tab from the events page. It was implemented separately and only the event name is sent to it. This allowed for development of the user interface in parallel with other aspects of the infrastructure.

\begin{figure}[htbp]
	\caption{\label{fig:tweetlist}
	Tweet list user interface in the new infrastructure.
	}
    \begin{center}
	\includegraphics[width=150mm]{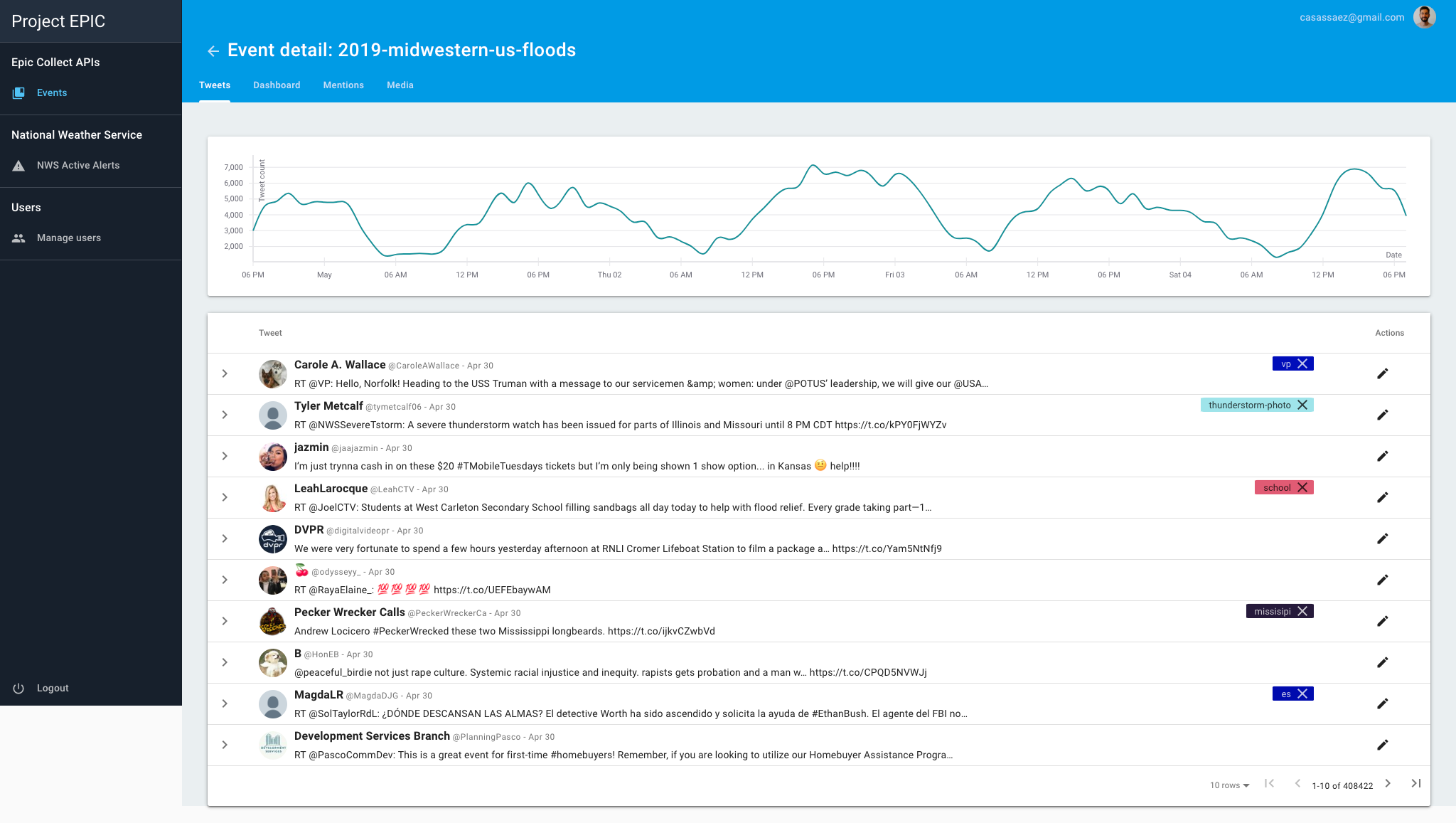}
    \end{center}
\label{xfigDiagram}
\end{figure}

To explore events, an interface was created similar to the one described for EPIC Analyze \cite{barrenechea2015getting}. There is a list of tweets with a timeline visualization above for time slicing the data set (see Figure \ref{fig:tweetlist}). Each tweet can be expanded to explore all the metadata associated with it, similarly to the functionality in EPIC Analyze. In addition, thanks to its composability, the user interface can also integrate external APIs easily. An example of this is the incorporation of  National Weather Service alerts. Using their public REST API that provides all current alerts information, I created a visualization of the active alerts. This way, the user interface could act as good integration with an external API, avoiding inserting new dependencies in the back-end.

To allow more technical analysts to perform deeper analysis, the system also points to the internal BigQuery table in Google Cloud Console. This allows analysts to explore the data set in a more interactive way using Google BigQuery directly.

Finally, in order to simplify interaction with each internal API, an ingress gateway was created. This portal aggregates all APIs under the same IP allowing the frontend to work as if it is a single API. In reality, each API is independent of each other. This gateway works by establishing rules for forwarding traffic to the corresponding API. This ingress is created by Kubernetes and associated with a static external IP, to avoid IP changes between cluster restarts. I also added an SSL certificate to the gateway using Kubernetes managed certificate service. This helped secure the application. 

\subsubsection{User interface diagram}

\begin{figure}[htbp]
	\caption{\label{fig:frontdiag}
	EPIC Dashboard interface diagram within React
	}
    \begin{center}
	\includegraphics[width=150mm]{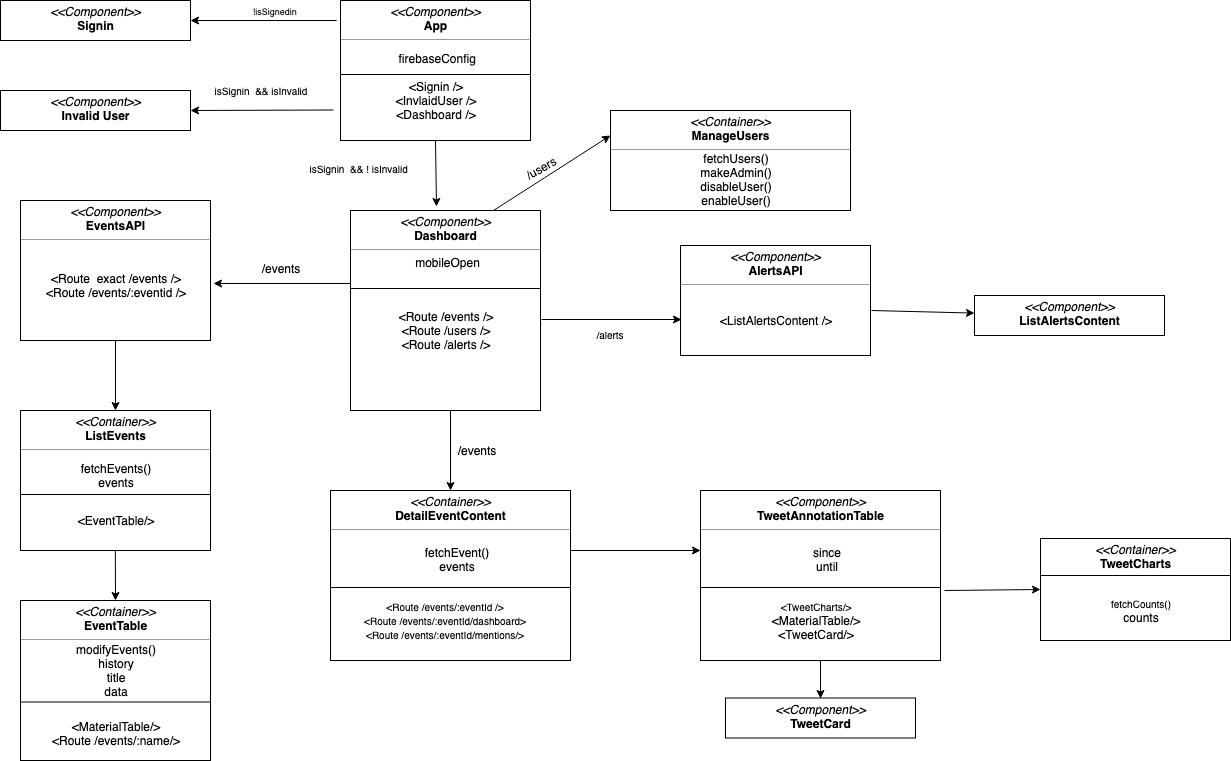}
    \end{center}
\label{xfigDiagram}
\end{figure}

See Figure \ref{fig:frontdiag} for the class diagram depicting the redux states and actions and their associated react components. I have segregated components into two types:

\begin{itemize}
    \item \textbf{Containers}: Containers are react components that are connected to redux i.e. use the “connect” component from the react-router-dom package.
    \item \textbf{Components}: These are often referred to as dumb components or simply components. They render the data based on the properties that are passed from their containers.
\end{itemize}

Class diagrams are not exactly built for react-redux classes. We have modified the classical class diagram by adopting the following conventions:

\begin{itemize}
    \item The class can be a container or a component. This is depicted in the first line within the $<<$ $>>$ notation.
    \item Just below the same notation, we have the class name specified. 
    \item The first block contains the variables from mapStateToProps and mapDispatchToProps. we distinguish between the action and the state variable in the following way: we append parenthesis with action names and the state variables are retained as is.
\end{itemize}

\subsubsection{Authorization/Authentication}

Since most of the back-end systems of my infrastructure are Google products (Google cloud, Google storage, etc.), it only made sense to continue relying on Google for authentication. I started to look into Google Firebase for the security and authentication of the system. It provides me with React UI components out of the box, thereby reducing development time. For sign up and sign in, I decided to use Google as the authorization mechanism. Given that the University of Colorado Boulder provides a personal Google account to all students and affiliates, I know that any Project EPIC analyst will have access to a Google account. 

Once a login is performed, Firebase issues an OAuth 2.0 authorization in the form of a JWT token. This token can be modified from the back-end to include arbitrary key-value pairs. In this case, I use this storage to decide whether a user can access the application. A simple key is stored internally on demand to give access to a user. This is synchronized using Google Cloud Identity Platform and Firebase. 

To manage this part, I designed the Auth API. This service queries the Google Cloud Identity Platform to list any users that tried to sign in with a Google account. In addition, the service checks whether a user has the specific key-value in their authentication permissions. This is returned in the form of JSON. Another endpoint allows for any signed-in user with authorization to revoke or enable access for any other user. This is a simple approach to user control and can fail for a large group of users. However, Project EPIC does not have a large number of internal users.

For authorization, the front-end sends the JWT token with each request to an API. This token is checked from the API using a Java library specific to Project EPIC. If the token is invalid or it does not have a good key-value, then the service does not allow the operation to proceed and returns a non-authorized error. This authorization process is injected into the application using the DropWizard authorization library. In addition to simplifying the development of the APIs, the authorization library allows quick disabling through a simple boolean. This feature allows testing APIs without a valid token and then requiring the token when they are deployed in production.

\chapter{Evaluation}

To evaluate my work, I will compare my new crisis informatics infrastructure with the previous infrastructure consisting of EPIC Collect and EPIC Analyze. My goal will be to evaluate my ability to replicate most of the previous infrastructure's features using cloud-based technologies. I will also evaluate whether my prototype is more extensible, scalable, and reliable.

\section{Data Collection and Persistence}

The main difference in data collection is the separation of concerns in separate services within the new infrastructure. The collection and filtering steps are separated into two different services. To update the keywords being collected, the new infrastructure uses a similar approach to the previous one. This time though, the collection service checks a local file which is updated by Kubernetes. The file is updated every time that the collection needs to be restarted by the Events API.

On the persistence front, I switched from Cassandra to files in an object storage system. This is better for the long term as there's no need for a database to be running to analyze the data. The primary reason for this change is the existence of unstructured scalable object analysis services provided by cloud services. The new system can still do data analysis yet make use of a simple persistence mechanism like file storage greatly simplifying the entire system and making it much more accessible to new Project EPIC developers in the future.

\subsection{Extensibility}

Thanks to the separation of stages in the collection pipeline, I can work on expanding the infrastructure's filter capabilities separately from the service that connects to the Twitter streaming API. In addition, thanks to the fact that I use Kafka to pass messages around, I can add additional real-time processing services that feed on the incoming stream of tweets. This can be done independently of the collection pipeline which makes it easier than the previous infrastructure to extend. 

Thanks to using a simple format for data storage, I can also add new analysis techniques on top of the collection service without having to worry about overwhelming a database. This also makes the process of data analysis simpler as it no longer requires having to manually copy data from our collection cluster to the analysis cluster. The data is always in the same place. In the future, new analysis software can be used to process all of the data. Since the system is storing the data in raw JSON format, it will be straightforward to connect new software with the existing data.

\subsection{Reliability}

Adding an external message passing software (Kafka) to use as a buffer increases reliability. If by any chance the filter service were to fail, Kafka will have a copy of all of the messages in its queue, so the filter service just needs to poll the buffer from Kafka and process those messages again. This ensures that the infrastructure does not lose messages due to filtering errors. 

With respect to data storage, Google Cloud Storage's contract ensures our data will be durable and available 99.9\% of the time. Thanks to this service guarantee, I avoid having to set manually any process for data redundancy or replication. With the previous infrastructure, even if there was data duplication, the data durability depended on the maintenance of our campus cluster and since Project EPIC had to have people to maintain our cluster, the project was in danger of losing the students who knew how to perform that maintainence. As students graduated, this risk became true and the cluster moved to a state where it was not being properly maintained. Due to the low maintenance caused by these changes, it became uncertain if all of Project EPIC's data would remain safely accessible. To mitigate this risk, I downloaded all of that data in early 2019 and stored it in Google Cloud Storage as a fail-safe. Thanks to this, all data sets for Project EPIC are stored in a similar pattern and are all stored in Google Cloud Storage and now data storage reliability is no longer a concern. 

\subsection{Scalability}

On the data collection aspect, the infrastructure can now scale horizontally different pieces of the pipeline individually. Kubernetes uses its horizontal scaling service for automatically scaling filters when  services have maxed their use of CPU resources. To manage workload partitioning between each instance of a filter service, the system relies on Kafka topic partitions, which uses internally a round-robin mechanism to assign each message to a partition. This enables a fast partition of data between nodes when scaling. In this fashion, the throughput of the ingestion pipeline can be increased as needed automatically. 

Compared to the previous infrastructure, the only way to scale the ingestion pipeline was by adding more CPUs to the server that was executing it. This type of scalability, known as vertical scalability, even if effective, can prove more expensive than horizontal scaling. CPU intensive machines tend to be more specialized machines that require a more specific setup, and therefore, tend to be more expensive to buy and maintain. In addition, such changes are expensive and so scaling in this fashion never occurred with the previous infrastructure.

Finally, to store the data, the system no longer has a potential bottleneck with the database as all of it's data are stored as files. This allows for a significant increase in the ability to store messages in parallel as multiple writes can occur to different files in the cloud storage system.

\section{Data Analysis}

With respect to data analysis, the main difference is the usage of BigQuery instead of the combination of Solr, Pig, abd Hive. BigQuery accepts SQL queries for the data sets without the need for using a traditional database. In addition, maintenance costs can be reduced thanks to not having to maintain the cluster that was running Solr, Pig, and Hive.

Google BigQuery can run more powerful filter queries than the ones we had in Cassandra. It can also replicate most of the capabilities available when using Solr in the old infrastructure, as my prototype can express EPIC Analyze queries as SQL in the new infrastructure. 

For the pagination and exploration of data sets, I was able to replicate similar results with the ones in EPIC Analyze by using filenames as the index to support pagination. Thanks to the index, I can also do quick time slicing of the data set similar to what is available in EPIC Analyze.

Finally, using Google Cloud Dataproc I have been able to replicate and indeed significantly outperform the capabilities of the simple job framework that was built for EPIC Analyze.

\subsection{Extensibility}

As mentioned before, thanks to having the collected data stored in a raw format, new analysis techniques can be added to the system with relative ease. In addition, I do not need to worry about causing a bottleneck on the database that could affect the collection pipeline. 

Thanks to Google Cloud Dataproc workflows, any developer can add new analysis jobs that provide new insights after collecting an event. New Spark jobs that export insights to Google Cloud Storage can be added. Paired with a new service, any developer can expose the results outside and further improve the user interface. This process has been proved that it can be done separately, an example of such is the mentions API discussed previously.

\subsection{Reliability}

BigQuery is reliable based on the service contract that Google Cloud provides. Thanks to making this service external, I reduced once again our maintenance costs, as I do not need to maintain a cluster for Hive or Pig as we did in the old infrastructure. This increases  system reliability as I am outsourcing this part to a third party. 

In that same way, having a single source for the data helps to avoid problems of loading data between the collection system and the analysis system as was done in the old infrastructure. This increases reliability in the long term as it is reducing the number of steps that need to be performed before an analyst can analyze a data set. 

\subsection{Scalability}

Thanks to BigQuery being a server-less component from Google Cloud, it works by scaling up for our query to match the size of the data. This is done internally by Google. For example, I can get the top 50 users who tweeted in a data set with 32 million tweets in under 2 minutes. However, compared to Solr, I can not perform filter queries as fast, the system is giving up performance from an index to gain performance on overall interactive queries. I could have used internal tables from BigQuery for increased performance, but given the number of queries launched to the data set over the life of a data set, it did not seem cost effective to do this yet.

Google Cloud Dataproc can also be configured to add more Spark nodes on demand. This allows adapting the post-processing batch workflow to be faster if needed. For the moment I left it at two nodes only, most data sets can be analyzed quickly under this configuration. 

Finally, thanks to the usage of a Load Balancer on top of the external services, I could scale up the external services using Kubernetes. This would allow to serve more data and adapt to peak usage of the user interface. This has not been implemented, as the current configuration should be more than enough for the average usage from Project EPIC.

\section{User interface}

The user interface is  the component that has accumulated the most changes in the new infrastructure. I added a new abstraction layer which is the service layer. This new layer separates the user interface from the data representation of the internal storage layers.

In addition, I switched to a more modern approach for designing web applications using a single page application framework. This helps to avoid bad practices of past web applications, separating views and controllers completely. The user interface is delegated to the browser, leaving the service layer in charge of abstracting the storage as a REST API. 

\subsection{Extensibility}

Thanks to React and how the dashboard app is structured, the user interface is easy to extend. New services can be integrated into the existing app by creating a new component to work as the user interface. Each component is in charge of managing its own API access, avoiding coupling between components.

To avoid retrieving data over and over, Redux is used to share state between components. Which means that new components using data which is already specified, can work without having to worry about how to retrieve the data from an API. 

Finally, the design is separated in separate css files from the main javascript file for each component. This separation allows for people to be more focused on designing interaction and visualization to work separately from the main logic. This can help in the future to create new data visualizations.

Compared to the previous user interfaces, the new user interface allows gathering under one roof all of the various steps needed to perform data collection and analysis. Previously an analyst had to use separate applications to collect data and analyze it. Now everything can be done under a single app. This helps to extend the capabilities from the user interface and to normalize the data across the analysis and collection sides of the system.

\subsection{Reliability}
The user interface is deployed separately from the services, which makes it more reliable. It is deployed to the Firebase CDN from Google. This allows the user interface to work, even if the internal services are having issues. This separation allows increasing reliability for the front-end, as before it was tied to the issues that may happen in Ruby on Rails applications.

\subsection{Scalability}

Views are served statically from a content delivery network as mentioned above. This means that the user interface is served with low latency across the planet and loaded fast into the browser. Compared to the previous infrastructure, this is a huge leap forward. Previously, the system was tied to the same server that was accessing the data internally. 

\chapter{Results}

After checking the new infrastructure and comparing it to the old version, I am going to explore now if I can answer the questions from my problem statement. 

First, in regards to maintainability, my proposed infrastructure simplifies the overall complexity of the system by outsourcing many parts to cloud services. Thanks to that, the system is not required to run on a large cluster of machines that Project EPIC has to maintain. The system goes from a set of six machines that are really specific to the infrastructure to four machines that run on a Kubernetes cluster.  In addition, my use of microservices separates system responsibilities into different components. This approach creates smaller services as compared to the monolithic components of the old infrastructure. Reducing the responsibility of each service makes the system easier to maintain. This is due to developers not having to understand the whole system at once and being able to work on single services individually. In addition, thanks to Kubernetes, developers do not need a system administrator to deploy their code; they just need to understand how containerization works. Once they have wrapped their microservice into a container, it can easily be deployed and plugged into all of the reliability and scalability features that Kubernetes provides. 

With respect to scalability, thanks to using BigQuery, the analysis pipeline scales better than before. This is due to this service being based on server-less cloud infrastructure. On the services side, with Kubernetes, services can scale horizontally using its horizontal auto-scaler. This is especially useful for the ingestion pipeline, as it allows for ingestion to adapt to demand. On the front-end, I can not even compare as both approaches are different. The new infrastructure though should be able to hold higher throughput thanks to the detachment of views and controllers. 

For reliability, Kubernetes provides abstractions to ensure the recovery of services automatically. Thanks to this and the microservice architecture, the system can be upgraded without bringing it down. There have been a few Kubernetes upgrades during development without much of a delay or error in the service provided. In addition, with Kubernetes rolling upgrades, the system first tests any new version of a service before switching incoming requests to it. In addition, the system switches from placing trust in Project EPIC being able to maintain hardware over time to a company that is focused on doing just that. 

Finally, this infrastructure has proved already that it can enable multiple extensions being developed in parallel. It is the case for the mentions API discussed above. The developer of that service was able to develop the extension fully on its own with ease. This proves that group work can be done on top of the current infrastructure in parallel, enabling easier collaboration between different researchers at Project EPIC, as each collaborator can work on their extension for the project with ease.

\chapter{Related Work}

There has been significant research into big data analysis and crisis informatics. I now present some related work to my thesis in the fields of crisis informatics and designing large data set analysis systems.  As mentioned above this work is an extension on top of the work that Project EPIC researchers have invested since 2009 in EPIC Collect and EPIC Analyze. 

\section{Data Collection and Storage for Crisis Informatics}

At Project EPIC there has been several papers describing how to collect tweets at scale. These have been related to a system known as EPIC Collect and are described in the earliest chapters of this thesis. In \cite{anderson2011design}, an early infrastructure based on MySQL to store data is proposed. This infrastructure is built using Spring to be scalable verically. Later in \cite{schram2012mysql} there is a switch from MySQL to NoSQL (Cassandra). This is due to the discovery of a bottleneck in the storage layer. However, the only way to scale this system remains vertically scalability as EPIC Collect needs a powerful machine to run its threads. A similar approach was exposed in \cite{kumar2011tweettracker}, without getting in depth on how the system was actually implemented.

The microservice approach in Project EPIC was proposed in my undergraduate thesis \cite{casas2017big}. The collection pipeline is pretty similar, changing to object storage instead of Cassandra. A similar approach is adopted by \cite{khaleq2018cloud}, the main difference being that their approach only allowed for a single event to be collected at a time. 

On the filtering side, most work has been done to collect using keywords. \cite{kejriwal2019pipeline} and \cite{khaleq2018cloud} propose different ways to automatically classify tweets as relevant for the event using machine learning. Other systems proposed suggest the usage of user timelines for a more contextual data set \cite{anderson2019incorporating}. For this most recent paper, I participated by developing microservices for the collection pipeline as well.

Other work has been using collection after an event using services like GNIP \cite{ashktorab2014tweedr}. It is important to note that due to recent changes in the Twitter API usage policy, these have become more expensive to use or simply unavailable. For this reason, live collection is important to keep costs down, even if analysis are only done after the event.

\section{Data Analysis for Crisis Informatics}

Project EPIC has done a lot of research on supporting analysts on their work to analyze large data sets. MongoDB was proposed in \cite{anderson2013architectural}, even though it was acknowledged that it had its limitations and the queries were slow to resolve. EPIC Analyze \cite{anderson2015design,barrenechea2015getting} addressed some of these issues by switching to an integration between Cassandra and Solr. 

In \cite{oussalah2013software} an approach for geospatial analysis is proposed. This work still limits the exploration capabilities of the analyst over the data set. Compared to the proposed system in this thesis, the new infrastructure expands to allow querying as a more generic usage while avoiding falling into complex systems difficult to maintain like the one proposed in \cite{oussalah2013software}. This approach was heavily influenced by the design of EPIC Analyze \cite{barrenechea2015getting}.

Other approaches have been to provide Spark on top of Cassandra stored tweets \cite{casas2017big}. However, this limits the fields to those defined on ingestion and makes the system fail if the schema ever changes. In addition, it has been found that Cassandra can act as a bottleneck when trying to use it for ingestion and analysis at the same time.

Finally, in \cite{madhavilatha2016streaming} it is suggested to use interactive programmatic approaches to streaming collections and their analysis. This work inspired me to open the usage of BigQuery for analysts to use, as sometime programmatic and flexible interfaces are required for more specialized analysts to be more productive. In this case, I do not try to reinvent the wheel and just allow the analyst to see the data and work with it using Google Cloud and BigQuery.

\chapter{Future work}

The main goal behind the infrastructure redesign has been extensibility. Therefore there are a lot of options to improve on extensibility in the future. At the beginning of the project, a set of features to develop was decided with other members from the software engineering group at Project EPIC. Some of the features were not implemented for my thesis. Others have just appeared after working on the project for a while.

An aspect that should be explored further is adding new methods for events. In this infrastructure, the only operations allowed for an event in the Events API are create, read and update. To be specific, the only update allowed is to start or stop the collection. Delete is not allowed since it would destroy data that could prove valuable. The design also does not allow to update the keywords since that may be confusing. If a keyword is removed from an event, does this mean that all the collected data must be removed if they were collected only for that keyword? I acknowledge that further research is needed as what should be done for each of the possible update scenarios.

More work could be done to extend the capabilities of filtering from the collection pipeline. Right now, an analyst can only specify whether a keyword should be in a tweet to collect. However, work can be done to add more rules for filtering. An option could be to avoid saving retweets. Another one could be to avoid saving tweets that mention a certain user. This can be extended through the filtering service as well as the Event API.

The Filter API can also be extended by adding new fields to query from. Using BigQuery, the system could be brought to a similar standing than the one available with Solr on the old infrastructure. Some of the proposed filters in previous Project EPIC are to filter by tweet author or to filter by whether it is a retweet or not. 

Another aspect to explore would be to explore filtered data sets for further work. An option to do so would be using BigQuery data sets. This feature could allow having multiple filtered data sets associated with an event, allowing analysts to work with them separately. One of the main difficulties would be to design a good user interface to compose these queries. Given that BigQuery allows for SQL queries, it would also be interesting to be able to obtain the actual query executed to use it in the BigQuery console. 

Another task would be to add filters for annotations in conjunction with tweets. This is a difficult modeling task, as annotations live in PostgreSQL and tweets live in the storage layer. 

It would also be interesting to add better support for media in data sets. An option could be to add automatic tagging of media using the content of the image and automatic tagging using convolutional neural networks. This would allow analysts to find images that they are interested in their research. 

Something more to explore is using Dataflow instead of Dataproc with Spark. Dataflow is a managed analysis service from Google Cloud. It may be worth switching over to avoid having Spark related issues with batch jobs. Dataflow also has workflows so it should be a seemingly straightforward transition. 

Another interesting extension would be to explore the pre-computation of indexes to explore tweets sorted by other fields than creation time. This would need further research as to whether it can prove useful to actually have it in the user interface as it is already available with BigQuery. If needed, it would be important to check whether it is better to use Spark on batch like the mentions API or if it is better to just use BigQuery to compute the ordered data on demand. 

Finally, it would be interesting to use automatic scoring for relevance. This could be done with machine learning. This score could be used to narrow down the data set to relevant tweets. This is an ambitious goal as it is complex. One way to do it would be using topic classification and allowing the user to filter using topics.

\chapter{Discussion}

After my work during my undergraduate thesis to explore containerization and container orchestrated systems, I was able to understand the advantages of using cloud infrastructures. With my thesis, I pointed at all the advantages of using microservices for collecting Twitter data. I also suggested building upon the prototype I did to create a new infrastructure for Project EPIC. This is fully realized in this current work. Here, I used the work done in my undergraduate thesis to expand from the collection pipeline and also supply the analysis interface. Thanks to my experience I have obtained through these years of research at Project EPIC, I have been able to design and fix errors I tackled in the previous iteration. In addition, I have been able to understand better the requirements from external and internal collaborators for this system.

I have also learned how to manage a team of students to develop this project. I believe it is a great experience and it has been really great to work alongside other people. I feel like some times research groups tend to work too separately and collaboration gets relegated as a small part. The results of this current work are such thanks to the active collaboration and management from the group of students I worked with. I would strongly encourage any future developments to incorporate a lead and more developers as this project did. 

Another topic of discussion is whether storing this data in external servers, like Google’s in our case, is a good ethical decision. Relying on external actors for research can lead to impartial results in favor of those that provide them. Given the latest scandals for bad usage of data from various companies in the tech industry, I believe that it is the duty for non-profit research to investigate and keep these companies accountable. It is in society’s best interest to have research on how private companies use our data to benefit themselves. This research can only come from impartial external parties, like universities. Relying on them to conduct research can prove to influence the process of accountability from universities. I believe it is important to power software engineering research in universities for this same reason. If universities fall behind from the industry, then we will only have ad-driven research which may not be the best benefit for society as a whole as we have seen in the latest years.

\chapter{Conclusion}
\OnePageChapter

Cloud infrastructures are changing the software engineering world by providing new alternatives to traditional architecture. This allows for new software architecture practices to be born. It is a change of perspective from traditional architectures. The more interesting part is the growth of server-less infrastructures, which are an evolution from container orchestrated infrastructures. Server-less services on the cloud will change how systems are designed in the future. New software engineering practices will be born from this new abstraction layer to provide future developers with new tools to design and build good software systems.

My research proves that there is a lot of potential within this world for software engineers. Research needs to be done in order to check if more use cases can benefit from the use of these tools as I did in my approach. 

In conclusion, I have been able to show that it is possible to reduce the maintenance cost of Project EPIC’s infrastructure, while replicating most of its features, in a cloud environment. Proving that cloud services can work in our favor to reduce costs, increasing reliability as well as scalability. While at the same time, it helps to reduce friction between collaborators to extend the system thanks to the microservice approach and to the reduction of dependencies within the system.

\bibliographystyle{plain}	
\nocite{*}		
\bibliography{refs}		

\appendix
\chapter{Code}	

All the code and deployment instructions are available online on opensource repositories in GitHub. The code is separated into 2 repositories. One for all infrastructure and backend and one for the user interface. All code is under version control on git. 

\section{Infrastructure and service code}

\textit{https://github.com/Project-EPIC/epic-infra}

Each service is organized into their corresponding folder. Inside a folder there's the code, the service documentation and instructions to upload the docker image to docker hub.

This repository also includes the Dataproc workflow definition. There you can add new analysis tasks to execute once an event has stopped collection. The Dataproc workflow needs to be overwritten and commited to maintain the file in the repository updated with the one deployed. There's also a separate folder for each Spark job. Each contains a \textit{pom.xml} file that describes how to compile and package the job code into an executable jar file.

Then there's the \textit{kubernetes} folder. In this folder, there's all the YAML definitions needed to deploy the system into a Google Cloud Kubernetes cluster. 

\section{User interface code}

\textit{https://github.com/Project-EPIC/reactdashboard}

All react code is contained inside the repository. It's structured following the sidebar navigation. Each navigation link on the sidebar has a folder in components. The repository follows a traditional structure for a React application.

\end{document}